# Purely analytic composites: Relative variance contributions of indicators corresponding to *a priori* indicator weights

André Beauducel[a*] & Ned Kock[b]

[a] *Department of Psychology, University of Bonn, Bonn, Germany*
[b] *Division of International Business and Technology Studies, Texas A&M International University, Laredo TX, USA;*
ORCID: André Beauducel 0000-0002-9197-653X
Ned Kock 0000-0002-5791-5434

**Abstract**

Composites are often created to facilitate the work of decision-makers. Therefore, practical or theoretical considerations may lead to *a priori* weights of the indicators forming a composite. Composites that are created a weighted aggregates are not the result of data analysis and may therefore be termed ‘analytic composites’. However, it has already been shown that the variance contributions of indicators within analytic composites are affected by the indicator variance and indicator inter-correlations. In the present study purely analytic composites are proposed, having exactly the variance contribution of indicators within the composites that are *a priori* defined by the indicator weights. An example based on simulated data illustrates the difference between analytic composites and purely analytic composites. As an application area, we propose that purely analytic composites could be of interest in the exchange-traded fund. An R-script for the computation of purely analytic composites is given in the Appendix.

*corresponding author
André Beauducel, Department of Psychology, University of Bonn, Kaiser-Karl-Ring 9, 53111 Bonn, Germany; email: beauducel@uni-bonn.de

## Introduction

A comprehensive definition of composites has been proposed by Nardo et al. (2008, p. 13): 'A composite indicator is formed when individual indicators are compiled into a single index on the basis of an underlying model.'. A main advantage of composites is that they can summarize complex, multi-dimensional realities which may facilitate the work of decision-makers. A main disadvantage of composites is that they can send misleading policy messages if they are poorly constructed or misinterpreted (Nardo et al., 2008). Typically, composites are created by the intention of individuals to create a weighted or unweighted aggregate of indicators. This implies that composites depend on theoretical models, expectations or practical considerations but that they are not the result of the analysis of empirical data. These intentionally weighted composites are also referred to as analytic composites (Kock & Beauducel, 2026a). Analytic composites, i.e., composites defined as in Nardo et al. (2008), should not be confounded with latent variables (LVs) or principal components. Although there is no single definition of LVs (Bollen, 2002), it is clear that the identification of LVs and principal components is based on empirical data analysis, like, for example, principal component analysis, factor analysis (Mulaik, 2009) or structural equation modeling (Byrne, 2012).

Although LVs can be related to aggregated manifest variables or indicators, they are not identical to these aggregates. This becomes apparent from the expected value definition of LVs by Lord and Novick (1968), where LVs are conceived as the expected value of its indicators. This definition implies that an LV is the aggregation of an infinite number of its indicators. Accordingly, the aggregate of a finite number of indicators can be an estimator of the respective LV although it can never be identical to the LV. Moreover, composites formed according to an intentional compilation of indicators in the sense of Nardo et al. (2008) are not formed as estimates of LVs but just as summary indices for decision making. We acknowledge that other definitions of composites might be appropriate in other contexts (Bollen & Bauldry, 2011; Kock, 2021; Kock & Beauducel, 2026a, b). However, the focus of the present paper is on the abovementioned definition of Nardo et al. (2008) because this form of composites is used in several studies and by several decision-makers (Bandura, 2006), while serious issues with their interpretation occur (Oswald, et al., 2015). Another term discussed in this context are composite formative indicators (Bollen & Bauldry, 2011) and there might be reasons for focusing on formative aspects of assessment, especially in the context of feedback processes (Jönsson, 2020) or interventions. However, there are so many complex meanings of formative measurement that we follow Hardin and Marcoulides (2011) in not using this term in the present context.

Starting from the perspective that analytic composites are theoretically and/or practically motivated weighted aggregates of indicators, we propose a method that allows to realize the intended relative variance contribution of indicators to the analytic composite. First, we report previous results by Oswald et al. (2015) to show that unknown empirical aspects of indicators can affect the weights of indicators so that they can deviate from *a priori* weights. Second, we propose a method that allows to maintain the intended relative variance contribution of indicators within the composite regardless of the empirical properties of the indicators. Third, we present a demonstration of the method by means of a population simulation. Finally, we discuss purely analytic composites in the context of other forms of indicator aggregation.

**The effect of empirical properties of indicators on indicator weights**

The simplest way to investigate the effect of empirical data on the weights of indicators within a composite is to consider unit-weighted indicators. Oswald et al. (2015) provide an example of the erroneous interpretation of composites based on unit-weighted indicators. The example concerns the erroneous idea that standardizing and summing unit-weighted indicators means that each indicator contributes equally to the composite. This is false as follows from

$$Var(\sum_{i=1}^{p} w_i x_i) = \sum_{i=1}^{p} w_i^2 Var(x_i) + \sum_{i=1}^{p}\sum_{i^{'}=1}^{p} w_i w_{i^{'}} Cov(x_i, x_{i^{'}}),\ with\ \ i \neq i^{'}, \tag{1}$$

where and $x_i$ is a vector of random variables representing $p$ indicators, $w_i$ are the indicator weights, $'Var'$ denotes the variance of the indicators and $'Cov'$ denotes their covariance (see Lord & Novick, 1968, Eq. 4.3.17). When all $w_i = 1$ and all $w_i^2 Var(x_i) = 1$, the variance of the unit-weighted aggregate, is

$$Var(\sum_{i=1}^{p} x_i) = p + \sum_{i=1}^{p}\sum_{i^{'}=1}^{p} r(x_i, x_{i^{'}}),\ with\ \ i \neq i^{'}, \tag{2}$$

where "*r*" denotes the correlation. Accordingly, $Var(\sum_{i=1}^{p} x_i)$ depends on $r(x_i, x_{i^{'}})$ so that $Var(r(x_i, x_{i^{'}})) > 0$ implies unequal variance contributions of the unit-weighted indicators within a composite. Indicators with a larger positive correlation with other indicators will have larger variance contributions within the composite. The different variance contributions of unit-weighted, standardized indicators according to their inter-correlation should be acknowledged when composites are formed by means of unit-weighted. standardized indicators. Equation 2 has already been shown by Oswald et al. (2015, p. 191). Although the effect of inter-correlations on the weights is more obvious in Equation 2, it also occurs in Equation 1, allowing for different *a priori* weights $w_i$. This implies that the variance contributions of indicators within analytic composites are not purely defined by the weights that are explicitly given to the indicators. There are unknown implicit weights resulting from the structure of inter-correlations of the indicators. This calls for purely analytic composites, where the weights given to the indicators are not affected by the inter-correlations of the indicators.

**Purely analytic composites: When a priori given indicator weights correspond to the variance contributions of the indicators**

Iterative procedures allowing to approximate equal variance contributions of standardized, unit-weighted indicators within a composite have been proposed by Wilks (1938) and Dunnette and Hogatt (1957). Here, the following solution is obtained with matrix-algebra:

Let $\mathbf{z}$ be a vector of $p$ standardized indicators, arranged with $p$ columns and the population of cases as rows. This implies that $E(\mathbf{z}'\mathbf{z}) = \mathbf{R}$, where "$E$" denotes the expectancy and $\mathbf{R}$ the inter-correlation matrix of indicators. Then, the composite based on standardized, unit-weighted indicators is

$$\mathbf{c} = \mathbf{z1}(\mathbf{1}'\mathbf{R1})^{-1/2}, \quad (3)$$

where "**1**" is a $p \times 1$ unit-vector and $\mathbf{1}'\mathbf{R1}$ is the variance of the composite. Accordingly, pre-multiplication with $\mathbf{z}'$ yields

$$E(\mathbf{z}'\mathbf{c}) = \mathbf{R1}(\mathbf{1}'\mathbf{R1})^{-1/2}, \quad (4)$$

Equation 4 illustrates that the correlation of the composite with the indicators depends on **R**, which was already shown in the previous section. When the indicators $\mathbf{z}$ in Equation 3 are post-multiplied with the inverse of **R**, the standardized composite is

$$\mathbf{c} = \mathbf{zR}^{-1}\mathbf{1}(\mathbf{1}'\mathbf{R}^{-1}\mathbf{1})^{-1/2}. \quad (5)$$

where $\mathbf{1}'\mathbf{R}^{-1}\mathbf{1}$ is the variance of **c**. Pre-multiplication with $\mathbf{z}'$ yields

$$E(\mathbf{z}'\mathbf{c}) = \mathbf{1}(\mathbf{1}'\mathbf{R}^{-1}\mathbf{1})^{-1/2}. \quad (6)$$

As $\mathbf{1}'\mathbf{R}^{-1}\mathbf{1}$ a scalar, Equation 6 implies that all correlations of the composite with the indicators are equal.

Composites with any other *a priori* relative contributions can be obtained when the $p \times 1$ vector **W** is inserted instead of "**1**" into Equation 5. This yields purely analytic composites, which are defined by

$$\mathbf{c} = \mathbf{zR}^{-1}\mathbf{W}(\mathbf{W}'\mathbf{R}^{-1}\mathbf{W})^{-1/2}, \quad (7)$$

so that

$$E(\mathbf{z}'\mathbf{c}) = \mathbf{W}(\mathbf{W}'\mathbf{R}^{-1}\mathbf{W})^{-1/2}. \quad (8)$$

As $E(\mathbf{z}'\mathbf{c})$ is the correlation of the indicator with the composite, the relative variance contributions are given by $E(\mathbf{z}'\mathbf{c})^2$. If the relative contributions are intended to represent directly the relative amounts of indicator variance in the composite, the weights in **W** should be specified as square roots of the intended relative variances.

**Indicator inter-correlations with analytic composites and purely analytic composites**

An example illustrating the effect of the inter-correlation variability of $p$ = 5 unit-weighted indicators on the correlation of the indicators with analytic unit-weighted analytic composites and with unit-weighted purely analytic composites is given in Figure 1. In the first population, the standard deviation of the indicator inter-correlations is $sd(\rho)$ = 0.01 and in the sixth population it is $sd(\rho)$ = 0.23. Populations with larger standard deviations of indicator inter-correlations have a larger variability of the correlations of the indicators with the unit-weighted analytic composites whereas the standard deviations of indicator inter-correlations does not affect the variability of the inter-correlations of the indicators with the purely analytic composites (see Figure 1). Accordingly, each indicator has the same variance contribution to the respective purely analytic composite, whereas the indicators have different variance contributions to the respective analytic composites.

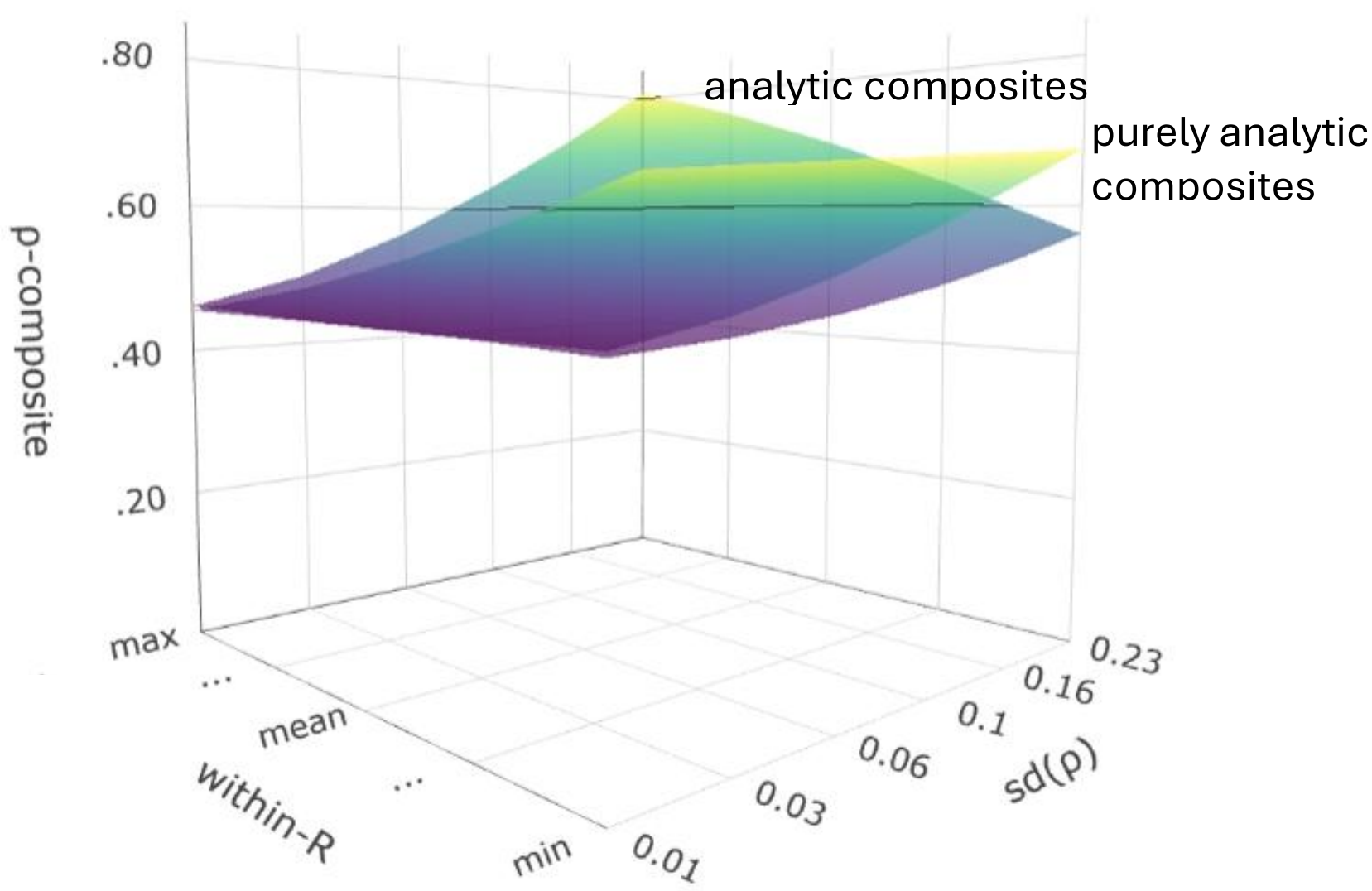

Figure 1. Standard deviations of indicator inter-correlations, $sd(\rho)$, minimum to maximum of indicator inter-correlations (within-R), and correlation of the indicators with the composite (ρ-composite) for unit-weighted analytic and unit-weighted purely analytic composites.

For the same population data used for the demonstration of the effect of indicator inter-correlation variability on the correlation of the indicators with unit-weighted analytic composites and unit-weighted purely analytic composites, a pattern of differential weights was also applied. The first three indicators had a unit-weight and the indicator four and five had a weight of two. This implies that the variance contributions of the indicators four and five to the composite should be four times larger than the variance contributions of the first three indicators.

The effect on the shared variance, i.e., the squared correlation of the indicators with the analytic composites, is presented in Figure 2A. According to the *a priori* weights, the shared variance of the indicators 4-5 with the composite should be four times larger than the shared variance of the indicators 1-3 with the composite. This occurs for $sd(\rho) = 0.01$, but this does not occur for $sd(\rho) = 0.23$ (see Figure 2A). For the purely analytic composites the shared variance is presented in Figure 2B. Here, the shared variance of indicators 4-5 with the composite is four times larger than the shared variance of the indicators 1-3 with the composite across all levels of $sd(\rho)$.

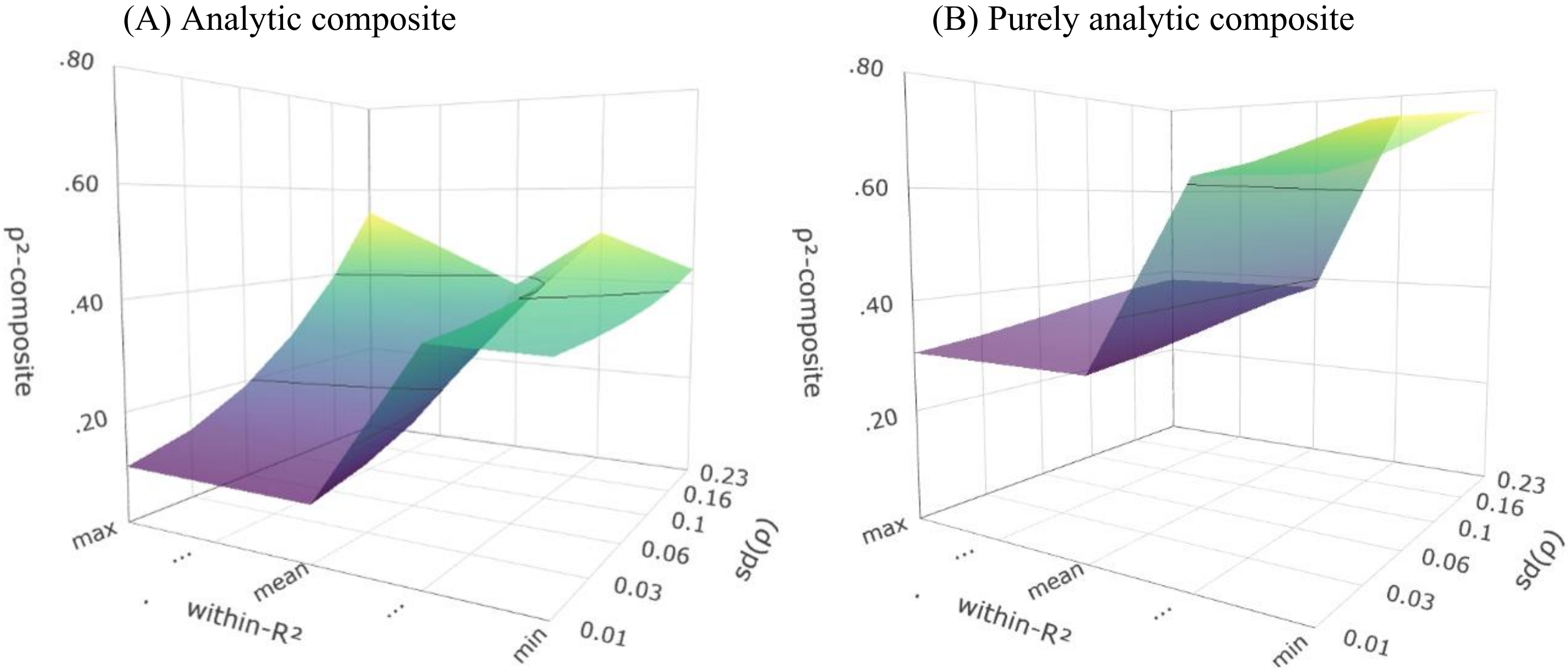

Figure 2. Standard deviations of indicator inter-correlations, $sd(\rho)$, minimum to maximum of squared indicator inter-correlations (within-$R^2$), and squared correlation of the indicators with the composite (shared variance: $\rho^2$-composite) for analytic composites (A) and for purely analytic composites (B).

**Area of application**

A promising application of purely analytic composites in the exchange-traded fund (ETF) domain lies in the construction of transparent, rules-based index methodologies that faithfully reflect investor-intended weightings. Traditional ETF indices often rely on weighted aggregates of underlying securities (e.g., market-cap weighting, factor tilts), yet these constructions can deviate from intended exposures because of covariance structures among assets. As the current paper shows, even when weights are explicitly assigned, the *actual variance contributions* of components depend on their inter-correlations and variances, introducing implicit weighting distortions. Purely analytic composites address this issue by ensuring that the variance contributions align exactly with the a priori weights, offering a theoretically cleaner foundation for ETF design.

In factor-based ETFs—such as those targeting value, momentum, or quality—this methodology could be particularly impactful. Factor ETFs typically combine multiple signals into a composite score, which then determines portfolio weights. However, correlations among signals (e.g., value and profitability) can unintentionally amplify or dampen certain exposures. Purely analytic composites eliminate this distortion by mathematically adjusting for inter-correlations, guaranteeing that each factor contributes to portfolio variance exactly as intended. This would enhance the interpretability of factor ETFs, aligning realized exposures more closely with their stated investment objectives.

Another important application is in multi-asset ETFs, where allocations are made across asset classes such as equities, bonds, commodities, and real estate. In conventional portfolio construction, even if an ETF targets, say, a 60/40 equity-bond split, the actual risk contribution may differ significantly due to changing correlations and volatilities. Purely analytic composites allow ETF designers to define weights in terms of desired variance contributions rather than nominal allocations. By doing so, the ETF could maintain stable risk budgeting across asset classes regardless of shifting market dynamics, improving consistency for investors seeking predictable risk profiles.

The approach is also highly relevant for thematic ETFs, which aggregate indicators related to themes such as clean energy, artificial intelligence, or emerging markets development. These themes are often operationalized through multiple indicators (e.g., revenue exposure, patent counts, ESG scores), combined into a composite ranking. Yet, as highlighted in this paper, analytic composites may misrepresent the intended importance of each indicator because of hidden empirical weights. Purely analytic composites ensure that thematic definitions remain faithful to their conceptual foundations, which is crucial for investor trust and product differentiation in a crowded ETF marketplace.

From a regulatory and disclosure perspective, purely analytic composites could improve transparency and accountability. Regulators increasingly scrutinize whether financial products deliver what they claim. If ETF providers can demonstrate that their weighting schemes correspond exactly to variance contributions—independent of underlying data peculiarities—they can offer stronger justification for their methodologies. This aligns with the paper's argument that when weights are theoretically justified, it is difficult to defend their contamination by empirical artifacts. Thus, purely analytic composites provide a defensible bridge between theory-driven design and empirical implementation.

Operationally, implementing purely analytic composites in ETFs would require integrating matrix-based transformations into index calculation engines. The paper provides a closed-form algebraic solution using the inverse of the correlation matrix to adjust indicator weights. While computationally more intensive than simple weighted averages, this is well within the capabilities of modern index providers. Moreover, the increasing use of smart beta and quantitative strategies suggests that the industry is already equipped to adopt such mathematically sophisticated approaches.

Finally, purely analytic composites could enable a new generation of "precision ETFs" that target exact risk contributions rather than approximate exposures. This would be particularly appealing to institutional investors and sophisticated retail participants who demand tighter control over portfolio characteristics. By eliminating unintended weighting distortions, these ETFs could offer more reliable performance attribution, better hedging properties, and clearer alignment with strategic asset allocation frameworks. In this sense, purely analytic composites are not just a technical refinement but a conceptual advancement that could reshape how ETFs are designed, marketed, and evaluated.

## Discussion

It has already been shown by Oswald et al. (2015) that composites formed by weighted aggregation of standardized indicators do not necessarily result in composites with relative variance contributions of the indicators that correspond to the initial composite weights. For unit-weighted indicators, iterative procedures to eliminate the divergence of indicator weights and indicator variance contributions have been proposed by Wilks (1938) as well as Dunnette and Hogatt (1957). In the present study, purely analytic components are computed by an algebraic method that can be applied to realize any *a priori* weights of standardized indicators as relative variance contributions of indicators within a composite. An illustration of the method based on simulated data is presented, exchange-traded funds (ETFs) are discussed as an area of application, and an R-script for the computation of purely analytic composites is given in the Appendix.

Form a broader perspective, purely analytic composites eliminate the effects of empirical properties of indicators on the resulting composites. Therefore, they are of interest when the theoretical or practical reasons for *a priori* weights of the indicators are most important. Moreover, purely analytic composites avoid unexpected confounding of empirical weights resulting from indicator variance and indicator inter-correlations with *a priori* weights. If composite weights are justified by *a priori* considerations, it is hard to conceive a justification for confounding the *a priori* weights with empirical weights. Therefore, purely analytic composites may be of interest when transparent theoretically defined weights are most important.

**References**

Bandura R. (2006), *A Survey of Composite Indices Measuring Country Performance: 2006 Update, United Nations Development Programme* – Office of Development Studies, available at http://www.thenewpublicfinance.org/background/Measuring%20country%20performance_nov2006%20update.pdf

Bollen, K. A. (2002). Latent variables in psychology and the social sciences. *Annual Review of Psychology, 53(1)*, 605–634. https://doi.org/10.1146/annurev.psych.53.100901.135239

Bollen, K. A., & Bauldry, S. (2011). Three Cs in measurement models: Causal indicators, composite indicators, and covariates. *Psychological Methods*, 16(1), 265–284.

Byrne, B. M. (2012). *Structural equation modeling with Mplus: basic concepts, applications, and programming.* Multivariate applications series New York: Routledge.

Dunnette, M. D., & Hoggatt, A. C. (1957). Deriving a composite score from several measures of the same attribute. *Educational and Psychological Measurement, 17*, 423–434. https://doi.org/10.1177/001316445701700309

Hardin, A., & Marcoulides, G. A. (2011). A commentary on the use of formative measurement. *Educational and Psychological Measurement, 71(5)*, 753–764. https://doi.org/10.1177/0013164411414270

Jönsson, A. (2020). Definitions of Formative Assessment Need to Make a Distinction Between a Psychometric Understanding of Assessment and "Evaluative Judgment". *Frontiers in Education, 5:2*. https://doi.org/10.3389/feduc.2020.00002

Kock, N. (2021). Common structural variation reduction in PLS-SEM: Replacement analytic composites and the one fourth rule. *Data Analysis Perspectives Journal*, 2(5), 1-6. Corpus ID: 254593970

Kock, N., & Beauducel, A. (2026a). On the causal priority of latent variables over indicators in SEM: Three main arguments. *Data Analysis Perspectives Journal, 7(2)*, 1-9.

Kock, N., & Beauducel, A. (2026b). On the causal priority of latent variables over indicators: Implications for SEM with composites and factors. *Data Analysis Perspectives Journal, 7(3)*, 1-9.

Lord, F. M. & Novick, M. R. (1968). *Statistical theories of mental test scores.* Addison-Wesley.

Mulaik S (2009). *Foundations of Factor Analysis, Second Edition* (Chapman & Hall/CRC Statistics in the Social and Behavioral Sciences): Chapman and Hall/CRC.

Oswald, F. L., Putka, D. J., & Ock, J. (2015). *Weight* a minute…What you see in a weighted composite is probably not what you get! In C. E. Lance & R. J. Vandenberg (Eds.), *More statistical and methodological myths and urban legends* (pp. 187–205). Routledge/Taylor & Francis Group.

Wilks, S.S. (1938). Weighting Systems for Linear Functions of Correlated Variables When There is no Dependent Variable. *Psychometrika, 3(1)*, 23-40. https://doi.org/10.1007/BF02287917

## Appendix

```
# r-script for computation of purely analytic composites


# Load necessary packages
# If the packages are not installed, run 'install.packages'
#install.packages('gdata')
library(gdata)
#install.packages('readr')
library(readr)


# Helper functions for frequently used matrix operations
Mdiag <- function(x) return(diag(diag(x)))
inv <- function(x) {x <- as.matrix(x)
if (det(x) <= 0.00001) {
  diag(x) <-  diag(x) + 0.000001
  return(solve(x))}
if (det(x) > 0) {return(solve(x))}}


# SET FILE DIRECTORY:
file_dir <- paste("C:/help/Ms/Reliability/",sep= "")

# ENTER THE NAME OF THE INDICATOR-SCORE-MATRIX:
# (txt-format, indicators are columns, scores are rows, use comma as separator, point for
# decimal places):
filnam_IND <- paste(file_dir,"Example_p=5_Indicators_n=100_cases.txt", sep="")
#filnam_IND <- paste(file_dir,"######.txt", sep="")

x <- read_delim(filnam_IND, delim = ",", quote = "\\\"", escape_double = FALSE,
                    trim_ws = TRUE, skip = 0, col_names = FALSE, na = c("") )
z <- scale(x)

# number of indicators
p <- ncol(x)
p

# number of cases
n <- nrow(x)
n

R <- t(z) %*% z /(n-1)
round(R, digits=2)


###################################### Filenames ################################
filnam_results                                                                           <-
       paste(file_dir,"purely_analytic_composites_p=",p,"_n=",n,"_results.txt", sep= "")
filnam_cor    <-    paste(file_dir,"purely_analytic_composites_p=",p,"_n=",n,"_indicator_inter-
       correlations.txt", sep= "")
filnam_scores                                                                            <-
       paste(file_dir,"purely_analytic_composites_p=",p,"_n=",n,"_composite_scores.txt",  sep=
       "")

# INSERT p A PRIORI WEIGHTS REPRESENTING THE WANTED RELATIVE VARIANCE CONTRTIBUTION OF
# INDICATORS WITHIN THE COMPOSITE:
W_var <- as.matrix(rbind(1,1,1,2,2))

#################################################################################



w <- W_var^0.5

anal_comp <-   z  %*% w  %*% (( t(w) %*% R %*% w ))^(-0.5)
cor_anal_comp <-   R  %*% w  %*% (( t(w) %*% R %*% w ))^(-0.5)
var_anal_comp <- cor_anal_comp^2
var_anal_comp_rel <- var_anal_comp/min(var_anal_comp)
help <- round(cbind(cor_anal_comp,var_anal_comp,var_anal_comp_rel), digits=3)
anal_comp_results <- cbind("Analytic comp.",help)
```

```
pure_anal_comp <-   z %*% inv(R)  %*% w  %*% (( t(w) %*% inv(R) %*% w ))^(-0.5)
cor_pure_anal_comp <-  w  %*% (( t(w) %*% inv(R) %*% w ))^(-0.5)
var_pure_anal_comp <- cor_pure_anal_comp^2
var_pure_anal_comp_rel <- var_pure_anal_comp/min(var_pure_anal_comp)
help <- round(cbind(cor_pure_anal_comp,var_pure_anal_comp,var_pure_anal_comp_rel), digits=3)
anal_pure_comp_results <- cbind("Purely analytic comp.",help)


results <- rbind(anal_comp_results,anal_pure_comp_results)
colnames(results) <- c("Composite", "cor. with comp.","var. within comp.","relative var. within
       comp.")
results


########################### Write results into files ###########################
options(warn=-1)
Message <- as.data.frame(paste("Compare analytic composites with purely analytic composites for
       n =",n, " and p =",p,":"))
write.fwf(Message, file=filnam_results,
          append = FALSE, quote = TRUE, sep = " ", na="", colnames=FALSE, rownames=FALSE,
       justify="left", width=100,
          eol="\n", qmethod=c("escape", "double"), scientific=FALSE)
write.fwf(as.data.frame(results), file=filnam_results,
          append = TRUE, quote = TRUE, sep = " ", na="", colnames=FALSE, rownames=FALSE,
       justify="left", width=22,
          eol="\n", qmethod=c("escape", "double"), scientific=FALSE)

write.fwf(round(R, digits=3), file=filnam_cor,
          append = FALSE, quote = TRUE, sep = " ", na="", colnames=FALSE, rownames=FALSE,
       justify="left", width=9,
          eol="\n", qmethod=c("escape", "double"), scientific=FALSE)

scores <- round(cbind(anal_comp, pure_anal_comp),digits=4)
colnames(scores) <- c("Analytic comp.", "Purely analytic comp.")

write.fwf(as.data.frame(scores), file=filnam_scores,
          append = FALSE, quote = TRUE, sep = " ", na="", colnames=TRUE, rownames=FALSE,
       justify="left", width=10,
          eol="\n", qmethod=c("escape", "double"), scientific=FALSE)


print("The weights and the correlations of indicators with composites are here:")
filnam_results
print("The inter-correlations of indicators are here:")
filnam_cor
print("The scores of the composites are here:")
filnam_scores
```